\documentclass[twocolumn,showpacs,preprintnumbers,amsmath,amssymb]{revtex4}

\usepackage{graphicx}% Include figure files
\usepackage{dcolumn}% Align table columns on decimal point
\usepackage{bm}% bold math

\begin{document}
%\preprint{Preprint}

\title{\emph{Ab initio} study of semiconducting carbon nanotubes adsorbed on the Si(100) surface:
 diameter- and registration-dependent atomic configurations and electronic properties}
 % study of semiconducting carbon nanotubes adsorbed on the
%Si(100) surface}
\author{Salvador Barraza-Lopez$^{1,3}$}
\email[Corresponding author: ]{barrazal@uiuc.edu} \affiliation{
1 Loomis Laboratory of Physics\\
2 Department of Electrical and Computer Engineering\\
3 Beckman Institute for Advanced Science and Technology\\
4 Materials Research Laboratory and Materials Computation Center\\
University of Illinois. Urbana, IL, 61801, USA}
\author{Peter M. Albrecht$^{2,3}$}
%\email{palbrech@uiuc.edu}
\affiliation{
1 Loomis Laboratory of Physics\\
2 Department of Electrical and Computer Engineering\\
3 Beckman Institute for Advanced Science and Technology\\
4 Materials Research Laboratory and Materials Computation Center\\
University of Illinois. Urbana, IL, 61801, USA}
\author{Nichols A. Romero$^{1,4}$}
\altaffiliation[Current address: ]{U.S. Army Research Laboratory,
Aberdeen Proving Ground, Maryland 21005-5066.}

 \affiliation{
1 Loomis Laboratory of Physics\\
2 Department of Electrical and Computer Engineering\\
3 Beckman Institute for Advanced Science and Technology\\
4 Materials Research Laboratory and Materials Computation Center\\
University of Illinois. Urbana, IL, 61801, USA}
\author{Karl Hess$^{1,2,3}$}
\affiliation{
1 Loomis Laboratory of Physics\\
2 Department of Electrical and Computer Engineering\\
3 Beckman Institute for Advanced Science and Technology\\
4 Materials Research Laboratory and Materials Computation Center\\
University of Illinois. Urbana, IL, 61801, USA}
\date{\today}

\begin{abstract}
 We present the first \emph{ab initio} study of semiconducting carbon
nanotubes adsorbed on the unpassivated Si(100) surface.
 Despite the usual gap underestimation in density functional theory, a dramatic reduction of the semiconducting gap for these hybrid
systems as compared with the electronic gaps of both their
isolated constitutive components has been found. This is caused by
the changes in the electronic structure as the surface
reconstructs due to the tube's proximity, the concomitant
electronic charge transfer from the nanotubes, and the band
hybridization with silicon and carbon states resulting in the
appearance of states within the energy gap of the formerly
isolated nanotube. Furthermore, it is determined that
semiconducting nanotubes exhibit weaker adsorption energies and
remain at a greater distance from the Si(100) surface as compared
to metallic nanotubes of similar diameter. This effect may be
useful for the solid-state separation of metallic and
semiconducting nanotubes.
\end{abstract}
\pacs{68.35.-p,68.43.-h,68.43.Bc,73.22.-f} \maketitle
\section{INTRODUCTION}
First-principles studies illustrating the effect of
technologically relevant semiconductor surfaces such as
InAs,\cite{Kim2004} GaAs\cite{Kim2005} and Si(100) on the
electronic properties of single-wall carbon nanotubes (SWNTs) have
been published recently. In the case of Si(100), the study was
focused on determining the lowest-energy structural configuration
and modifications to the electronic structure of a metallic (6,6)
SWNT as a result of its interaction with this surface. This
calculation was performed for a nanotube in the proximity to an
either clean\cite{Orellana2004_1,Orellana2004_2} or selectively
hydrogen-passivated\cite{Orellana05} Si(100) surface where the
nanotube axis was parallel to the trench between adjacent Si dimer
rows. Remarkably, there have been no studies of this system for
\emph{semiconducting} tubes, in different geometrical
configurations, nor on the dependence of the properties of the
hybrid system against nanotube diameter. These are relevant issues
to be addressed, as experimental techniques permitting the
ultra-clean deposition of SWNTs onto doped Si(100) and other
semiconductor surfaces in ultra-high vacuum (UHV) at room
temperature have been
reported.\cite{Albrecht2003,Albrecht2004,Ruppalt2004} As
contaminant-free atomistic manipulation becomes more feasible, the
promise of molecular systems with tunable electronic and
mechanical properties becomes a reality.

 It is clear that near-term applications for carbon nanotubes in electronic and opto-electronic devices
would benefit from their integration with conventional
semiconductor platforms such as Si or GaAs. In this direction,
rectifying carbon nanotube-silicon heterojunction arrays have been
demonstrated,\cite{Tzolov2004} and semiconductor heterostructures
of GaAs/AlAs and GaAs/MnAs have been used as electrical contacts
to individual SWNTs.\cite{Jensen2004}
 Recent experiments\cite{Su2000} showing the existence of two preferential
directions for the growth of SWNTs on silicon surfaces indicate
that there is a non-negligible interaction between the SWNT and
its substrate. One relevant issue which remains to be addressed is
the nature of the mechanical and electronic properties of the
semiconductor-nanotube interfaces.

The subject of this work is the interaction between the
unpassivated Si(100) surface and semiconducting carbon nanotubes
using density functional theory (DFT).\cite{hohenberg64} Those
studies are complemented with additional calculations on metallic
nanotubes of comparable diameter. We study nanotubes that are in
parallel or perpendicular orientations relative to the Si dimer
row direction. We have found a striking and somewhat
counter-intuitive reduction of the semiconducting gaps for these
hybrid systems which are composed of two semiconductors. The
nanotubes presented in this study have diameters between 5 and 12
\AA. This work was comprehensive in order to find robust
properties on these hybrid systems; in particular those
independent of diameter and chirality as experiments do not have
fine control of these variables yet.

Section \ref{sec:methods} discusses the theoretical approximations
employed and the structural configurations considered in this
study. In Section \ref{sec:results} we show the resulting atomic
configurations of the combined Si(100)-SWNT system, which have
marked trends depending on the original electronic character of
the SWNTs involved (metallic or semiconducting). Results from
Voronoi and Hirshfield population analysis\cite{Fonseca2003} are
also provided and show electronic charge being transferred from
the nanotubes to the silicon slab. The resulting band structures
at equilibrium, as well as the projected density of states (PDOS)
over the atomic species involved are reported also. The PDOS
indicates a clear hybridization of the bands between surface and
nanotube states. This hybridization is further confirmed by
depicting wavefunctions near the electronic gaps. Finally, we
present the change in energy prior to the force minimization
procedure, as the nanotubes are rotated about their axes in close
proximity to the Si(100) surface. Conclusions are presented in
Section \ref{sec:conclusions}.

\section{methods}\label{sec:methods}
Our calculations in the local density approximation\cite{Kohn65}
(LDA) were performed with the SIESTA code.\cite{Soler2002} The
exchange-correlation potential employed is the one parameterized
by Perdew-Zunger,\cite{pz} based on the Ceperley-Alder
data.\cite{ca} Core electrons are replaced by norm-conserving
Troullier-Martins pseudopotentials.\cite{tm} For greater
variational freedom, a double-$\zeta$ basis set for $s$ and $p$
orbitals, and a single-$\zeta$ basis set for $d$ orbitals was
constructed using the prescription of Junquera \emph{et
al.}\cite{Junquera01} To ensure the flexibility in our basis sets,
their parameters were optimized by means of the simplex algorithm
\cite{Pressrecipes}
 on graphite and diamond for the carbon basis, while the silicon
basis was optimized only in the diamond structure. As the hydrogen
atoms only served to passivate dangling bonds, the hydrogen basis
was not optimized.
 As shown in
Table \ref{tab:table1}, the lattice constants obtained from the
optimized bases compare well with previous theoretical estimates.
\begin{table}
\caption{\label{tab:table1}Lattice constants obtained with our
bases, to exemplify the flexibility of our basis set.}
\begin{ruledtabular}
\begin{tabular}{lcc}
Element&Theoretical estimate (\AA)&This work (\AA)\\
 \hline
 C (Diamond) &3.570${}^{\dagger}$  & 3.486\\
 C (Graphene)&$a$ 2.450${}^{\ddagger}$&$a$ 2.448\\
             &$c$ 6.500${}^{\ddagger}$&$c$ 6.516\\
 Si (Diamond)   &5.430${}^{\dagger}$&5.401
\end{tabular}
\end{ruledtabular}\\
\begin{flushleft}
$\dagger$Reference~\onlinecite{AshcroftMermin}.\\
$\ddagger$Reference~\onlinecite{Kganyago2003}. \end{flushleft}
\end{table}
 We use the p(2$\times$2) reconstruction of the Si(100) surface for calculations involving
semiconducting SWNTs. The gap obtained is of 0.235 eV. In all
cases a slab with six silicon monolayers with a height of 7.81
\AA~was employed. The bottommost layer is hydrogen(H)-passivated
forming a dihydride arrangement. The SWNTs will be placed in
proximity of the uppermost unpassivated layer. The area of the
unit cell is $L\times{}L$ with $L=7.639$ \AA{}. The vacuum region
in the vertical direction is fixed in order to provide at least
$10$ \AA{} separation between periodic images when the SWNTs are
in place.

Since we are interested in trends that could complement both
experimental work in this area and published theoretical results
on metallic SWNTs on this surface the SWNTs chosen satisfy the
following criteria:
\begin{enumerate}
\item{}They are semiconducting.
\item{}Their diameter is of the order of 10 \AA{}.
\item{}They are commensurate with the underlying surface.
\end{enumerate}
While ($n$,$n$) (metallic) SWNTs happen to be commensurate to
within 3\% with the underlying Si(100)
surface,\cite{Orellana2004_1} the shortest commensurate
semiconducting SWNT, of indexes (3$n$,$n$) --and $3n-n\neq3q$ with
$q$ integer-- would be 15.391 \AA{} long. The large unit cell
sizes involved in calculations have deterred researchers from
performing calculations in this system with semiconducting
nanotubes. Table \ref{tab:table2} lists the nanotubes studied,
their diameters, lengths as well as their DFT gaps. Tubes with
chiral indexes (2$n$,$n$) are 11.294 \AA{} long, so a supercell
constructed out of two nanotube unit cells is required in this
case to meet the surface's supercell length. Thus the slab
supercells required for calculations involving both the surface
and semiconducting nanotubes have surface areas equal to
$3L\times{}2L$ or $3L\times{}3L$, depending on the length of the
SWNT, its diameter and the surface reconstruction employed. This
choice leaves more than 10 \AA{} between nanotube images.
\begin{table}
\caption{\label{tab:table2}Carbon nanotubes with length in their
unit cell commensurate within 2\% to the Si(100) surface's
supercell. In bold, the respective values for the (6,6) SWNT
previously studied by Orellana \emph{et
al.}\cite{Orellana2004_1,Orellana2004_2,Orellana05}}
\begin{ruledtabular}
\begin{tabular}{cccc}
Nanotube&Diameter (\AA)&Length (\AA)&DFT Gap (eV)\\
 \hline
(6,2)&5.657&15.391&0.860\\
(8,4)&8.302&11.294&0.945\\
(12,4)&11.314&15.391&0.699\\
(9,3)&8.486&15.391&0.000\\
\bf{(6,6)}& \bf{8.153}&\bf{2.465}&0.000
\end{tabular}
\end{ruledtabular}
\end{table}
In order to test the accuracy of our calculations against
published results,\cite{Orellana2004_1,Orellana2004_2} benchmark
calculations on a chiral metallic SWNT interacting with c(4x2)
reconstructed Si(100) surface were performed. The surface gap in
this case is equal to 0.223 eV. The c(4$\times$2) reconstruction
minimizes the total energy the most and therefore is the best
candidate for the surface reconstruction.\cite{Healy2001} The
metallic (9,3) SWNT is comparable in diameter to the (6,6) SWNT
studied extensively by Orellana \emph{et
al.}\cite{Orellana2004_1,Orellana2004_2,Orellana05} Due to the
surface reconstruction used and the requirement for the nanotube
images to be separated by at least 10 \AA{} this required an even
larger, $4L\times{}2L$ area for the underlying surface. This
benchmark calculation is also used to assess the effect of
chirality and the exchange-correlation approximation; LDA in this
work and GGA in Refs.
\onlinecite{Orellana2004_1,Orellana2004_2,Orellana05}.

\section{Results}\label{sec:results}
\begin{figure*}
\scalebox{0.8}{\includegraphics{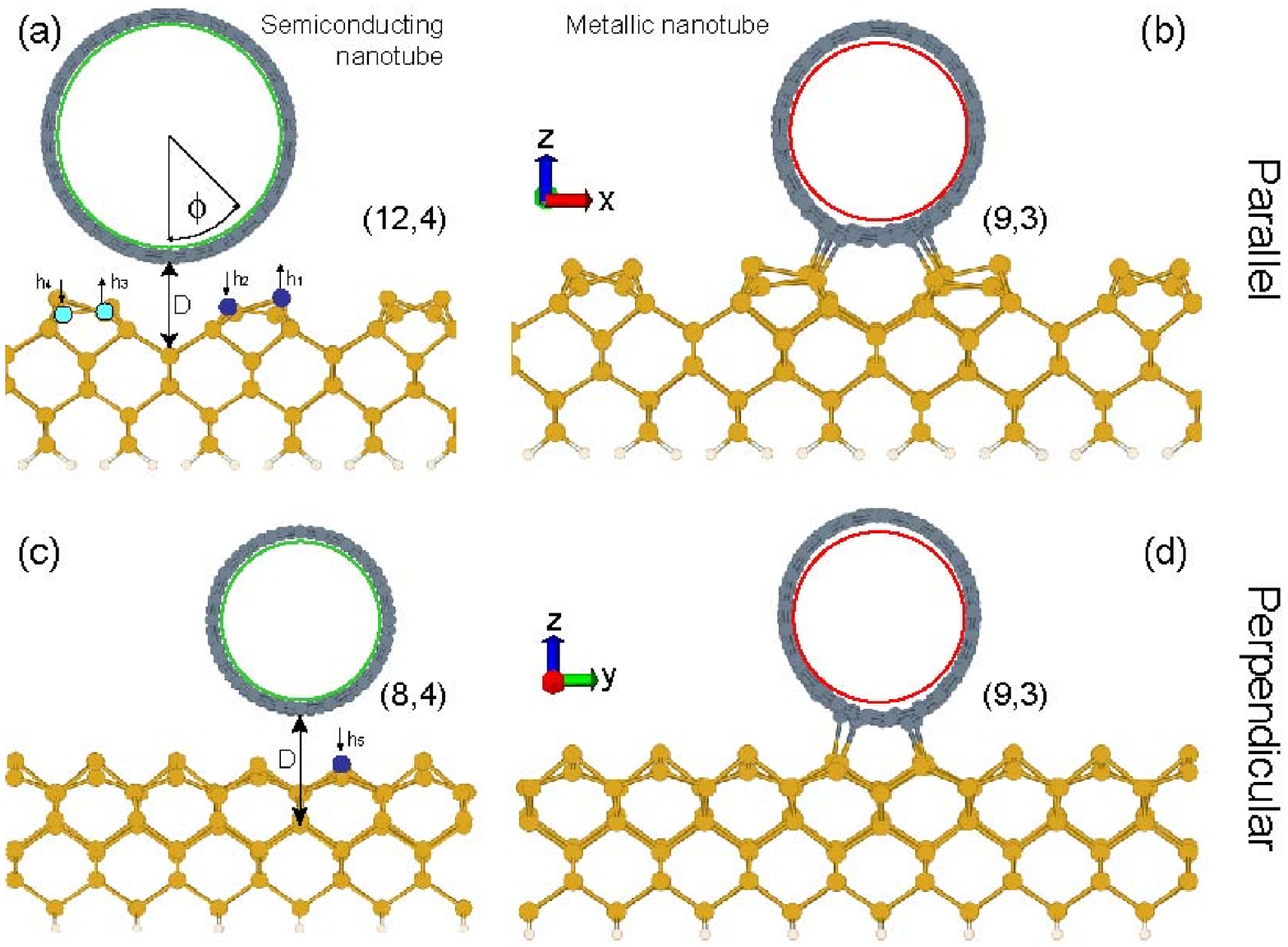}} \caption{\label{fig:Fig1}
(Color online) SWNTs on the Si(100) surface in two configurations:
\emph{parallel} over the dimer trench, (a)-(b) and
\emph{perpendicular} to the Si(100) surface dimer trench, (c)-(d).
$D$ is the distance from the bottom of the SWNT to the bottom of
the Si(100) surface dimer trench obtained in the total energy
minimization. The vertical displacements $\langle{}h_i\rangle$
involved in the surface reconfiguration and $\phi$, the angle of
rotation of the SWNT about its axis, are also indicated. An
evident `squashing' effect --an elongation along the z-direction--
can be observed for the (9,3) nanotube but it was not present in
any semiconducting nanotube. Note that for chiral tubes, a
rotation $\phi$ about their axis and a displacement $\Delta l$
(not shown) along the nanotube's axis, are linearly dependent. To
minimize the cell size, a p(2$\times$2) surface reconstruction was
employed when placing semiconducting nanotubes, and the
c(4$\times$2) surface reconstruction for the (9,3) nanotube.}
\end{figure*}
\begin{table}
\caption{\label{tab:tableConf} Averaged displacement (over the
unit cell) in the z-direction (\AA) of silicon atoms closest to
the nanotube; c.f. Fig.~\ref{fig:Fig1}. $\langle
h_1\rangle-\langle h_4\rangle$ refer to the parallel configuration
and $\langle h_5\rangle$ to the tubes in the perpendicular
configuration.}
\begin{ruledtabular}
\begin{tabular}{crrrr}
Nanotube:&(6,2)&(8,4)&(12,4)&(9,3)\\
\hline
$\langle$h$_1\rangle$&0.15&0.11&0.03&0.14\\
$\langle$h$_2\rangle$&$-$0.18&$-$0.29&$-$0.41&$-$0.44\\
$\langle$h$_3\rangle$&0.20&0.15&0.09&0.43\\
$\langle$h$_4\rangle$&$-$0.05&$-$0.06&$-$0.04&0.00\\
\hline $\langle$h$_5\rangle$&$-$0.13&$-$0.26&$-$0.16&$-$0.36\\
\end{tabular}
\end{ruledtabular}
\end{table}
Figure \ref{fig:Fig1} shows hybrid structures with maximum
residual forces less than 0.02 eV/\AA. The relevant geometric
parameters are the minimum distance $D$ between the SWNT and the
dimer trench, the vertical atomic displacements
$\langle{}h_i\rangle$ resulting from the reconfiguration of the
silicon atoms closest to the SWNT, and the azimuthal rotation
angle $\phi$. We want to emphasize the shorter bond lengths
between the carbon and silicon atoms when the metallic tubes are
involved, as well as a marked distortion of the nanotube,
emphasized by the inner (red) circle. In comparison,
semiconducting tubes are in equilibrium farther away from the
surface and do not change their circular profile, as evidenced by
the (green) inner circle perfectly matching the nanotubes'
perimeters. To contrast with published
work,\cite{Orellana2004_1,Orellana2004_2} due to chirality the
metallic nanotube in either configuration studied showed only
seven bonds of varying lengths (2.03--2.27~\AA{} in the parallel
configuration and 2.02--2.15~\AA{} in the perpendicular
configuration). For the semiconducting nanotubes, the shortest
distance between carbon and silicon atoms turned out to be
2.62~\AA{} and 2.87~\AA{} for the (6,2) tube in the parallel and
perpendicular configuration, respectively. All the semiconducting
tubes with larger diameters had their carbon atoms more than
2.84~\AA{} away from the closest surface silicon atom and in most
cases more than 1~\AA{} above the distance in the C--Si bond of
silicon carbide. This is indicative of a weaker bonding for
semiconducting nanotubes adsorbed on the Si(100) surface. For
chiral nanotubes, the optimal geometrical configuration can not be
known \emph{a priori}, but it has to come out after a sweeping
through all angles $\phi$; this angular dependence is not as
pronounced for semiconducting SWNTs. A detailed discussion
follows.

\subsection{Optimized atomic configurations}\label{sec:structures}

  An equivalent plane-wave cutoff of 200 Ry was used to calculate
the charge density on the real-space grid. To obtain the optimal
configuration of these systems, we used a grid
 with a single $k$-point (the $\Gamma$--point), except for the (6,2) nanotube, where a
2$\times$2$\times$1 Monkhorst-Pack\cite{mp76} (MP) grid was
already employed for obtaining the relaxed structures. The
adsorption energies were then calculated with a
2$\times$2$\times$1 MP grid.

\subsubsection{Distance from the SWNTs to the silicon surface and surface reconfiguration}
\begin{figure}
\scalebox{1.0}{\includegraphics{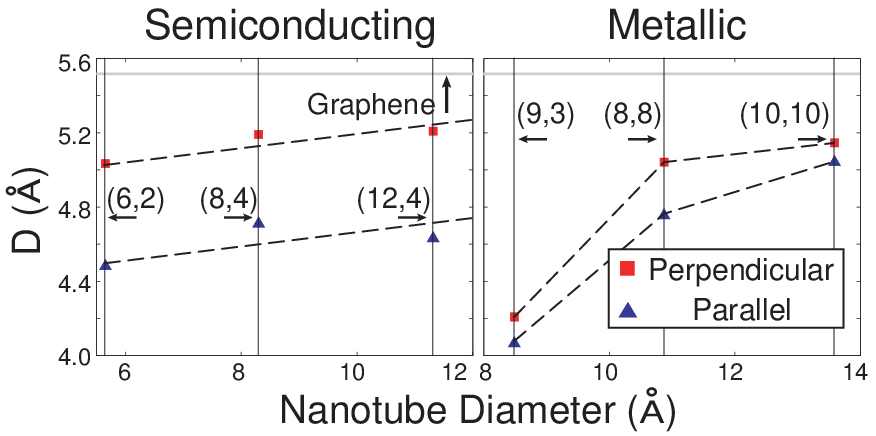}}
\caption{\label{fig:Fig2}(Color online) Distance $D$ from the
bottom of the dimer trench to the lowest carbon atom as a function
of nanotube diameter for fixed $\phi$. Semiconducting SWNTs
parallel to the dimer trench will be closer to the underlying
surface by about 0.5 \AA{} as compared to semiconducting SWNTs in
the perpendicular configuration for this diameter range. Notice a
more pronounced diameter dependence and an overall smaller
distance to the surface for metallic SWNTs. Dashed lines are drawn
as a guide to the eye to facilitate the visualization of trends.}
\end{figure}
\begin{figure}
\scalebox{0.4}{\includegraphics{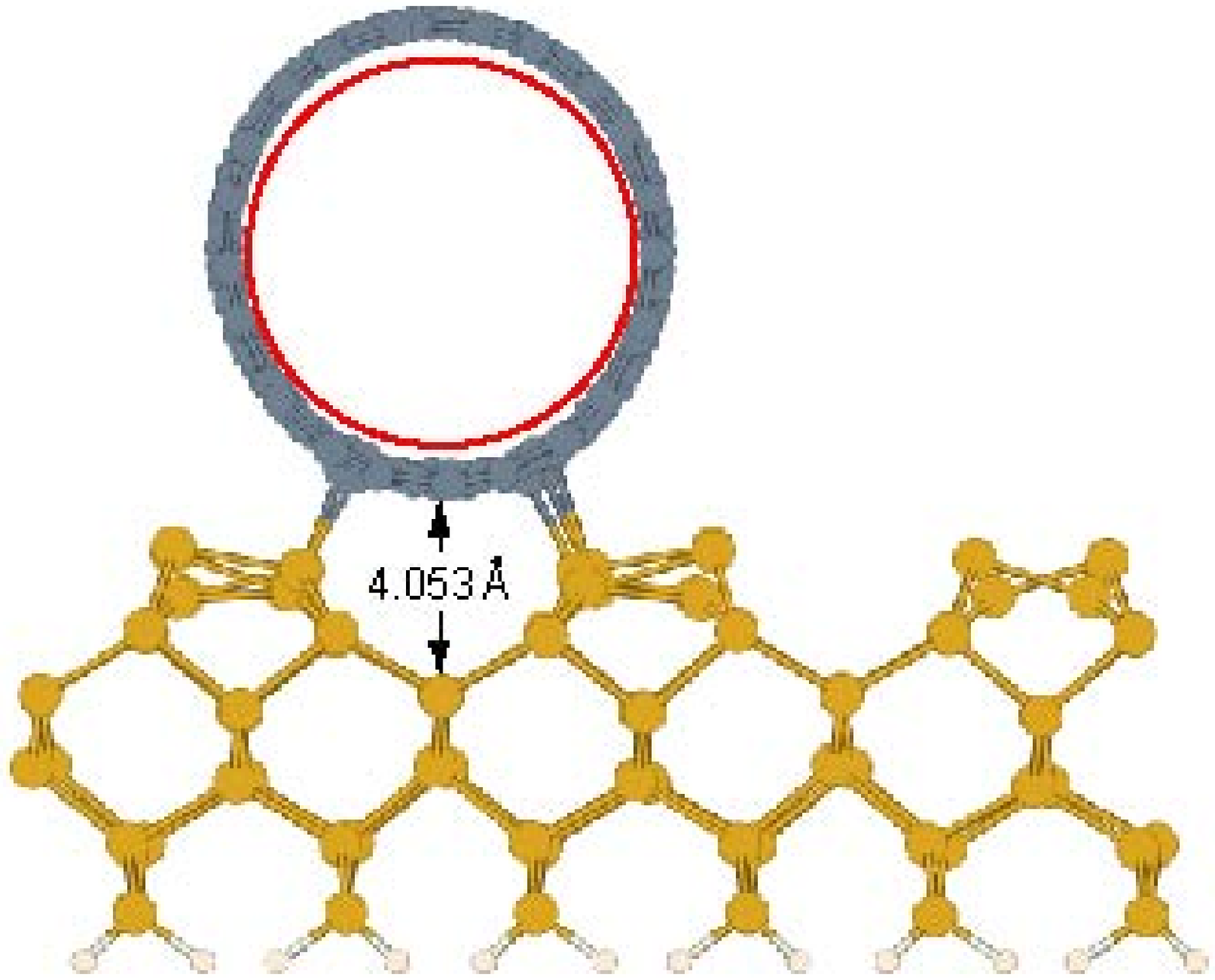}}
\caption{\label{fig:Fig93} (Color online) The properties of the
hybrid system do not depend on the underlying surface
reconstruction as evidenced by the (9,3) nanotube on the
p(2$\times$2) reconstructed surface. The distance between the
nanotube and surface is very close to that found for this nanotube
on the c(4$\times$2) reconstructed surface, see Fig.
\ref{fig:Fig2}. Notice also the vertical elongation appearing
again and the overall structural similarity to the structure
presented in Fig. \ref{fig:Fig1}.}
\end{figure}

Total energies for the hybrid SWNT-Si(100) system with the SWNTs
listed in Table \ref{tab:table2} were calculated as a function of
the distance from the nanotubes to the bottom of the the dimer
trench. The optimal distance $D$ is reached when the total energy
is at the global minimum. While holding the bottom hydrogen layer
fixed, the hybrid structures were relaxed by a conjugate-gradient
method to minimize the residual forces until they no longer
exceeded 0.02 eV/\AA. We also calculated the optimal distance $D$
for (8,8) and (10,10) SWNTs for a fixed angle $\phi$ in order to
visualize and distinguish trends between semiconducting and
metallic SWNTs. No relaxation procedure was performed on the
latter tubes. Under relaxation those distances would decrease
slightly, but the trend indicated here will stand. It should also
be mentioned that the trend obtained in Fig.~\ref{fig:Fig2} for
metallic tubes remained the same when the c(4$\times$2)
reconstructed surface was replaced by a p(2$\times$2)
reconstructed surface. This means that the distinct trends
observed in Fig.~\ref{fig:Fig2} are due to the different
electronic character of the nanotubes involved, and \emph{not} to
the surface reconstruction employed. This is stressed in Fig.
\ref{fig:Fig93}, where the structural configuration of the (9,3)
tube and the p(2$\times$2) reconstructed surface with maximum
residual forces down to 0.02 eV/\AA{}, appears extremely similar
to the one found for this nanotube on the c(4$\times$2)
reconstructed surface.
 The diameter of the nanotube as well as its electronic
character (metallic, semiconducting) are the determining factors
on the relative distance between the nanotube and the surface. As
the nanotube diameter increases, the distance $D$ from the SWNTs
to the bottom of the dimer trench also increases asymptotically
towards 5.52 \AA, which is the limiting distance from the bottom
of the Si dimer trench to a graphene plane (nanotube diameter
$\to\infty$), indicated by the gray line in
Fig.~\ref{fig:Fig2}.\footnote{In order to obtain the separation
between the surface and the graphene plane, we constructed silicon
surface of area $4L\times{}L$. A graphene surface can be placed on
top in which the vector
$\mathbf{v}\equiv{}\mathbf{a}_1+\mathbf{a}_2$ is perpendicular to
the dimer trench, with a 3\% elongation of the graphene plane in
both perpendicular directions ($\mathbf{a}_1$ and $\mathbf{a}_2$
are the graphene lattice vectors). We then found a minimal energy
by varying the separation between the graphene plane and the
relaxed surface.} The results in Fig.~\ref{fig:Fig2} show that
metallic nanotubes exhibit a stronger diameter dependence on their
optimal distance to the surface's slab.\footnote{This dependence
of $D$ on the nanotube's diameter is overlooked in Reference
\onlinecite{Orellana2004_1}.} In previous work, the same distance
to the surface was used for the (6,6) SWNT and a tube of 100 \AA{}
in diameter, in an attempt to explain experimental results of
multiwalled SWNTs on \emph{H-passivated}
surfaces.\cite{Hertel1998} We speculate that the adsorption energy
of nanotubes on an H-passivated surface should be lower than the
one corresponding to an unpassivated surface, because the dangling
bonds of the silicon surface are hydrogen passivated. This effect
is worthy of further investigation, but is not included in this
present study. For the semiconducting nanotubes studied the
diameter dependence is not as marked. Yet, Fig.~\ref{fig:Fig2}
shows that a semiconducting SWNT in the perpendicular
configuration will be about 0.5~\AA{} farther away from the
silicon surface in comparison with its equilibrium distance in the
parallel configuration, in accordance with experimental
observation.\cite{Albrechtu}

 The electronic properties of the Si(100) surface are
closely related to its atomic configuration. Table
\ref{tab:tableConf} gives the average vertical displacements
$\langle{}h_i\rangle{}$ that the silicon atoms closer to the SWNT
undergo due to the proximity of the SWNT. The Si(100) surface in
the proximity of the SWNT is subject to a striking atomic
reconfiguration. For a given SWNT the amount of surface
reconfiguration is always greater in the parallel case. This is
consistent with the fact that tubes in the parallel configuration
stay closer to the surface (Fig.~\ref{fig:Fig2}) and therefore
interact more strongly than in the perpendicular case. We have to
stress here that after atomic relaxation the semiconducting
nanotubes undergo almost no structural change but remain very
stiff. This is not the case for the metallic nanotube, as
previously indicated.

\subsubsection{Relative strength of the C-Si bonds}
\begin{figure}
\scalebox{0.3}{\includegraphics{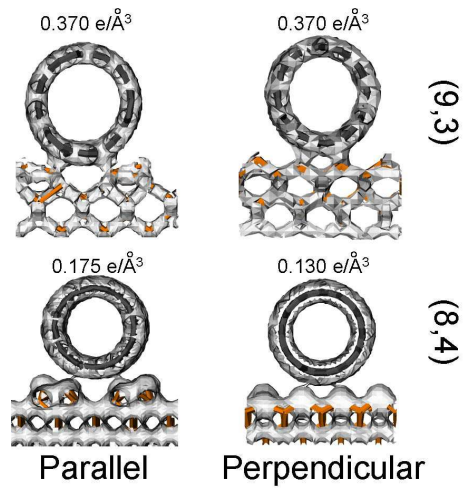}}
\caption{\label{fig:bonds}(Color online) Charge density isosurface
plots to visualize the relative strengths of C--Si bonds. Values
of the densities associated with each isosurface are shown. The
strongest bond occurs for the metallic SWNT.}
\end{figure}
With the goal of visually expressing the varying strengths of
bonds formed between carbon and silicon atoms for the metallic and
semiconducting SWNTs, we show in Fig.~\ref{fig:bonds} charge
density isosurfaces for the (9,3) and (8,4) SWNTs. Results for the
remaining nanotubes can be found in Table \ref{tab:tableDen}. Very
dense bonds can be seen for the (9,3) SWNT in either configuration
with the 0.37 $e$/\AA$^3$ isosurface. The charge density
isosurfaces show that the C--Si bonds are as strong as the Si--Si
and C--C bonds in the surface and the nanotube, respectively, as
can be inferred from the similar width.
\begin{table}
\caption{\label{tab:tableDen} Charge density ($e$/\AA$^3$) at the
bonds between nanotube carbon and surface silicon atoms. Notice
the smaller density for all studied semiconducting SWNTs in
comparison with the bond densities for the metallic nanotube.}
\begin{ruledtabular}
\begin{tabular}{lrrrr}
Nanotube:&(6,2)&(8,4)&(12,4)&(9,3)\\
\hline
Parallel&0.235&0.175&0.120&$>$0.370\\
Perpendicular&0.155&0.130&0.115&$>$0.370
\end{tabular}
\end{ruledtabular}
\end{table}
  In order to understand the varying strengths
 of the bonds and the role of chirality on bond formation, refer
to Fig.~\ref{fig:Fig1}b, depicting the (9,3) SWNT in
 the parallel configuration. In this figure, surface dimers are
 located in the ($x$,$z$) plane. For the shortest bond (2.03 \AA{}) the relative coordinates of the involved carbon and silicon
atoms are ($\Delta x$, $\Delta l$, $\Delta z$)=($-$1.00, $-$0.23,
1.75) \AA. The second closest carbon atom to a surface silicon
atom has relative coordinates equal to (0.99, 0.35, 1.76) \AA,
giving a bond length of 2.05 \AA. Although the cartesian
projections $|\Delta x|$ and $\Delta z$ are very similar in
 both cases, chirality breaks the mirror symmetry  with respect to
 a plane formed by the SWNT axis and the bottom of the dimer
 trench once all bonds are considered (recall that only seven --instead of eight-- bonds were formed), but it also shifts
 the carbon atoms by varying distances $\Delta l$, according to
 their positions in the SWNT unit cell. Contrast this to the more symmetric configurations previously studied for the (6,6) SWNT
 in which pairs of bonding atoms have the same $\Delta l$ and a similar value for $|\Delta x|$ for all bonds due to mirror symmetry.

 For the (8,4) SWNT we had to lower the density to 0.175
$e$/\AA$^3$ to identify bond formation in the parallel
configuration. Lower density through bonds will reflect the weaker
adsorption energies found for semiconducting SWNTs. Notice that
this isosurface does not form bonds in the perpendicular
configuration. It is at the lower 0.130 $e$/\AA$^3$ density when a
single isosurface connecting the nanotube and slab can be seen.
This is about a third of the density present in the strongest
C--Si bond for the metallic SWNT.
 In conclusion, bond formation will be weaker for semiconducting
 SWNTs, and the weakest bond occurs for semiconducting SWNTs in
 the perpendicular configuration. These results are
 consistent with all findings of previous subsections.

\subsubsection{Amount of charge transferred to the silicon slab}

The results from Voronoi and Hirshfield deformation density charge
analysis\cite{Fonseca2003} are summarized in Table
\ref{tab:tableVor}. Because of the way they are defined, the
results from those methods are independent of the numerical
orbital basis, and they are also more meaningful physically: The
Voronoi deformation density charge for a given atomic nucleus, for
instance, is the absolute charge defined in the volume defined by
all the points closer to that atomic nucleus than to any other
nuclei. In all instances, the charge is transferred from the
nanotube to the slab, and the magnitude of the charge transferred
is larger when the tubes are parallel and on top of the dimer
trench.
\begin{table}
\caption{\label{tab:tableVor}Amount of electronic charge per unit
length ($e$/\AA) transferred from the SWNT to the slab from the
Voronoi (Hirshfield) deformation density analysis.}
\begin{ruledtabular}
\begin{tabular}{ccc}
Nanotube&Parallel&Perpendicular\\
 \hline
(6,2)&+0.052 (+0.046)&+0.033 (+0.029)\\
(8,4)&+0.055 (+0.049)&+0.035 (+0.033)\\
(12,4)&+0.072 (+0.065)&+0.035 (+0.031)\\
(9,3)&+0.064 (+0.083)&+0.058 (+0.043)
\end{tabular}
\end{ruledtabular}
\end{table}

\subsection{\emph{Ab initio} electronic properties of the hybrid Si(100)-SWNT system}
\subsubsection{Band structures and densities of states}
\begin{figure*}
\scalebox{0.9}{\includegraphics{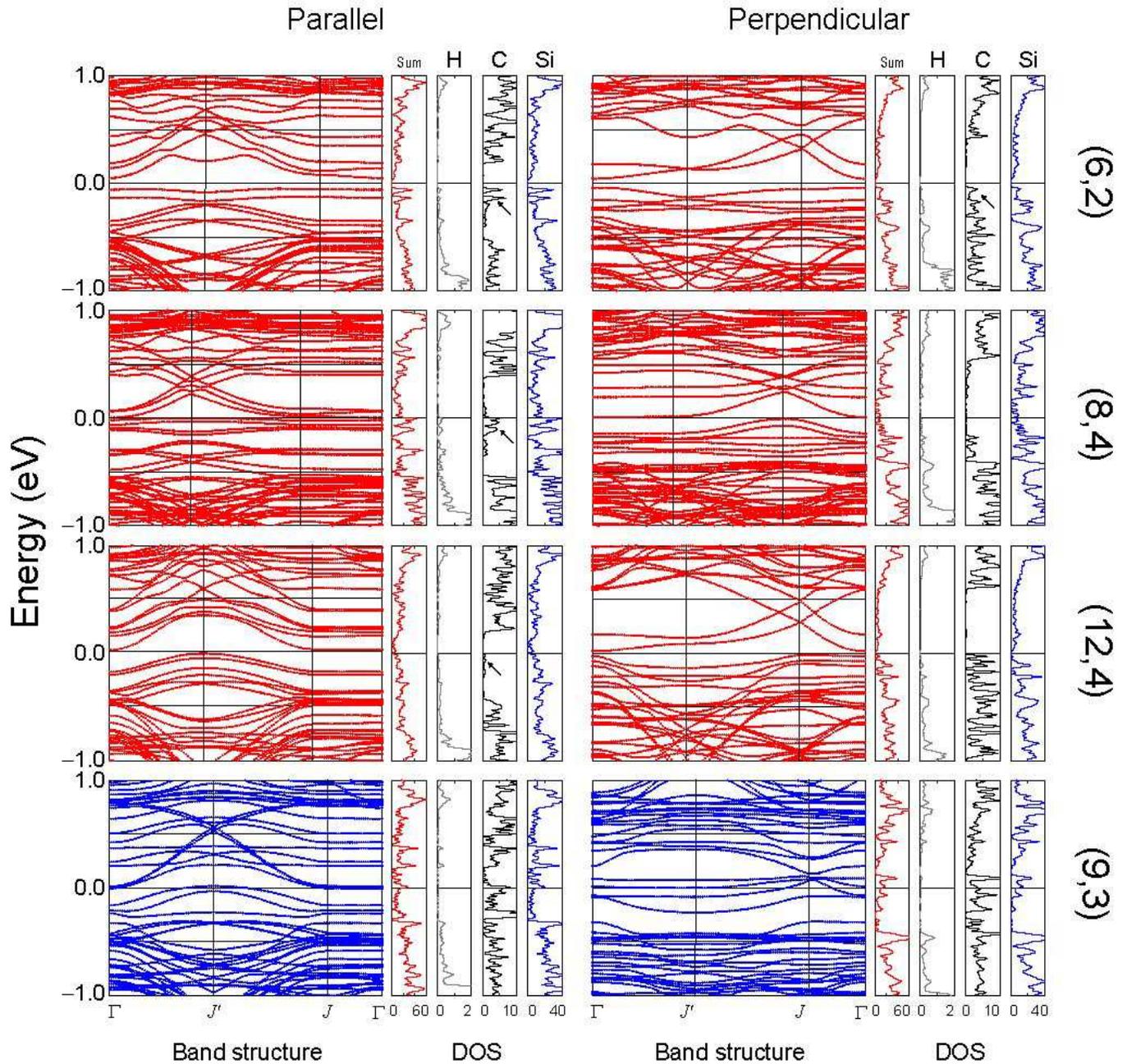}}
\caption{\label{fig:Figbands}(Color online) \emph{Ab initio} band
structures and projected densities of states for SWNTs in
different alignments with respect to the Si(100) surface. The
first column shows the band structures and PDOS when SWNTs are
aligned on top and parallel to the dimer trench, while the second
column depicts band structures when the nanotubes are
perpendicular to the dimer trench. Those results are obtained
after atomic relaxation was performed. The \emph{ab initio}
results show a drastic reduction of the semiconducting gap for the
hybrid system composed of semiconducting tubes and the Si(100)
surface and a high degree of band hybridization. The different
positions for the $J$, $J'$ points reflects the difference in size
of the unit cells considered. The band structures involving the
metallic nanotube (in blue) show no gap opening at the Fermi
level. The band structures are calculated with a
4$\times$4$\times$1 MP grid, while the PDOS was obtained with at
least a 12$\times$12$\times$1 MP grid. Arrows indicate the
contribution to the band structure from carbon atoms in the
vicinity of the Fermi level.}
\end{figure*}

\begin{figure*}
\scalebox{0.8}{\includegraphics{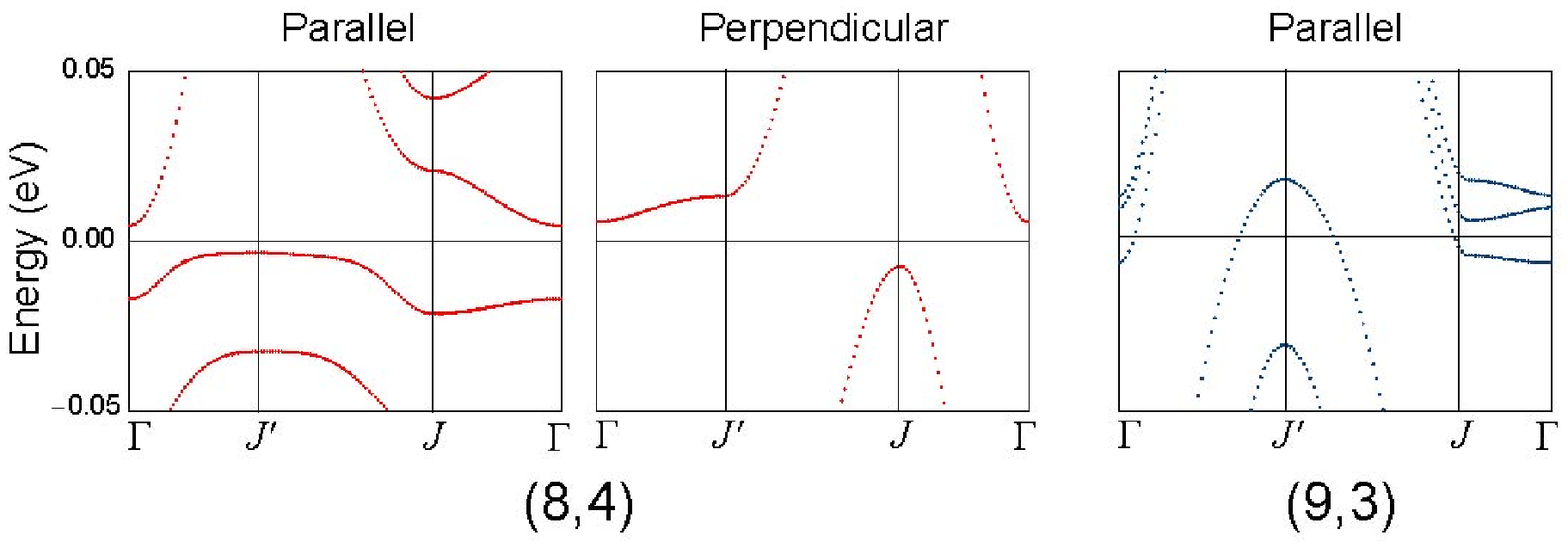}}
\caption{\label{fig:Zoom}(Color online) Band structure for the
(8,4) and (9,3) nanotubes around the Fermi energy to visualize gap
openings or lack thereof. We omit zooming into band structures for
which the gaps can be directly visualized from
Fig.~\ref{fig:Figbands}.}
\end{figure*}

In order to have reliable band structures a stringent force
relaxation was performed. As mentioned in Section
\ref{sec:structures}, the atomic positions were subject to a force
minimization procedure with a single $k$-point (the $\Gamma$
point), with the exception of the (6,2) SWNT, where a
2$\times$2$\times$1 MP grid was employed. The relaxed atomic
positions from this calculation were then used in computing the
band structures and PDOS. The band structures were obtained with a
4$\times{}$4$\times{}$1 MP grid, while the PDOS was computed from
at least a 12$\times{}$12$\times{}$1 MP $k$-point grid (for
example, the results for the (12,4) nanotube in the parallel
configuration were obtained with a 24$\times$12$\times$1 MP grid),
using the converged self-consistent charge density obtained from
the calculation with a 4$\times$4$\times$1 MP grid. The band
structures and PDOS are depicted in Fig.~\ref{fig:Figbands}. In
all the plots shown in this figure the SWNT was placed parallel to
the $\Gamma-J'$ direction. For the PDOS a Gaussian smearing of 10
meV half-height width and 300 sampling points in the ($-1,1$) eV
energy interval were employed. The Gaussian smearing is in most
cases smaller than the electronic gaps obtained from the band
structures.\footnote{The choice of Gaussian smearing is consequent
with the k-point sampling employed. In particular, a finer
Gaussian smearing requires a concurrent finer mesh and it is
beyond our computational capabilities.}  The PDOS is computed with
a finer mesh, independently from the band structure calculation
and indicates the contribution of each atomic species to the band
structure.
\begin{table}
\caption{\label{tab:tableGaps}Electronic gaps (in eV) for the
hybrid systems composed of semiconducting nanotubes on Si(100)
surface. We also include the (9,3) tube for completeness.}
\begin{ruledtabular}
\begin{tabular}{lrrrr}
Nanotube:&(6,2)&(8,4)&(12,4)&(9,3)\\
\hline
Parallel&0.099&0.008&0.033&0.000\\
Perpendicular&0.083&0.013&0.040&0.000
\end{tabular}
\end{ruledtabular}
\end{table}
 The two columns in Fig.~\ref{fig:Figbands} correspond to the
 relaxed hybrid system in either the parallel or
 perpendicular configurations. The most salient feature from Fig.~\ref{fig:Figbands}, as the band structures indicate, is
the dramatic reduction of the gap for the hybrid systems composed
of semiconducting nanotubes (with gaps of at least 699 meV when
isolated) and the Si(100) surface (with a gap bigger than 200
meV). The biggest gap found occurred for the system involving the
(6,2) nanotube, and is equal to 83 meV. Refer to Table
\ref{tab:tableGaps} for a comprehensive list of the electronic
gaps. The projected densities of states in Fig.~\ref{fig:Figbands}
indicate a contribution from the carbon orbitals to the band
structure at energy values where the gap of isolated tubes is
expected, highlighted by the arrows on the carbon PDOS (note the
different scales for each species in the PDOS). We believe this
reduction of the gap for the hybrid system with respect to the
gaps of its constituent subsystems will occur even when more
accurate quasiparticle (e.g. GW) calculations and experimental
data are in place. Notice that, despite the strong structural
modification observed for the metallic tube (Figs. \ref{fig:Fig1}b
and \ref{fig:Fig1}d), this is insufficient in this case to open a
gap in either configuration; refer particularly to
Fig.~\ref{fig:Zoom} where we zoomed the band structure about the
Fermi level in the parallel case, to better appreciate this fact.
 The reader might appreciate the existence of a single band
above the Fermi energy for the system involving the (8,4) SWNT in
the perpendicular configuration. The PDOS indicates a couple of
dips and the density at those dips is \emph{not equal} to zero.
The discrepancy here only comes due to the insufficient number of
$k$-points employed. This system has the largest number of atoms,
and we are certain a calculation with a finer mesh will result in
a flatter PDOS in the region of interest, but we did not pursue
further calculations due to their expensive nature. In the same
note, the computational cost involved in computing a PDOS for the
zoomed bands, which requires a finer smearing and a
correspondingly increased number of $k$-points, prevents us from
performing such calculation. Nevertheless we are certain of the
accuracy of the band structure, and therefore of the accuracy of
the electronic gaps for the hybrid system with a (8,4) tube
adsorbed. Experimentally, the gaps for the nanotubes could be
obtained accurately on the inert H-passivated substrate, prior to
hydrogen depassivation; the surface gap can be measured by a host
of experimental techniques. The dramatic reduction of the gap in
this hybrid system as well as the distinct structural trends for
semiconducting and metallic nanotubes are the main findings of our
investigations.

In Ref.~\onlinecite{Berber2006} a gap opening for the
small-diameter and achiral (5,5) nanotube was found in the
perpendicular configuration. Due to a lesser symmetry in chiral
tubes, which translates in longer bond lengths and a more complex
atomic reconstruction and band hybridization, this effect is not
seen in the (9,3) nanotube reported on this paper, and certainly
more systematic studies would be required to address such a rather
interesting effect; particularly with regards to knowing if it is
robust enough to be observed in other nanotubes. Our work
indicates that any semiconducting nanotube on this semiconductor
surface results in a hybrid system with a gap smaller from that of
the isolated constituent subsystems.

\subsubsection{Highest occupied, and lowest unoccupied electronic states for the Si(100)-SWNT system}
\begin{figure*}
\scalebox{0.7}{\includegraphics{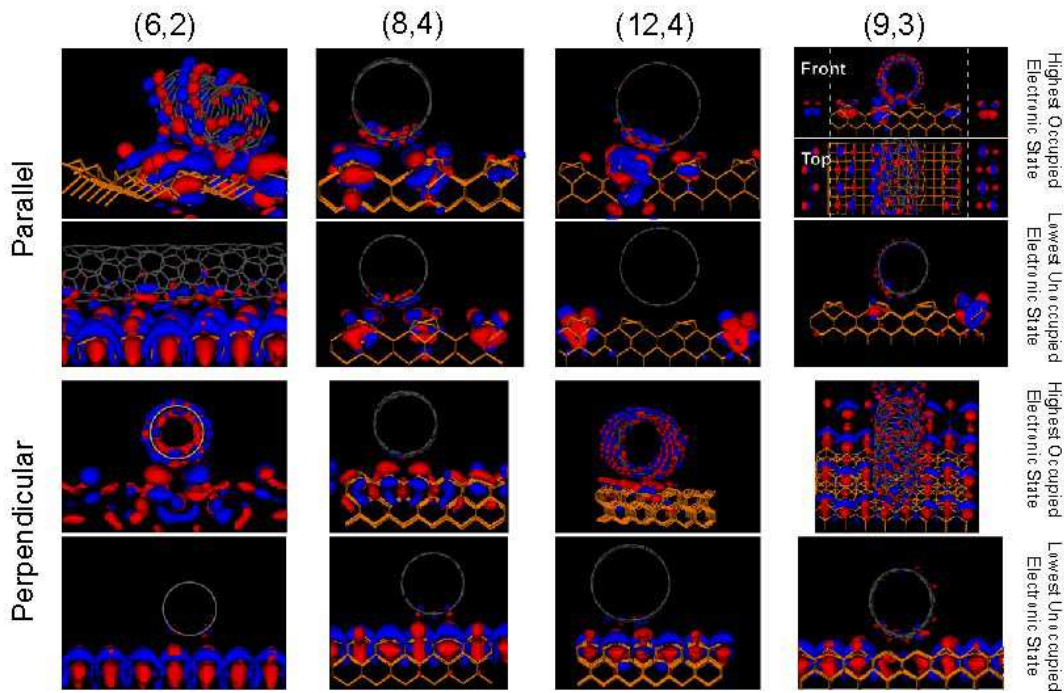}}
\caption{\label{fig:WFS}(Color online) Wavefunctions depicting
$\Gamma$-point states right below (highest occupied electronic
state) and above (lowest unoccupied electronic state) the Fermi
energy. The hydrogen bottom layer is not shown.}
\end{figure*}
Wavefunctions for electronic states at the $\Gamma$-point that are
just below and above the Fermi energy --the ($\Gamma$-point)
Highest Occupied and Lowest Unoccupied Electronic States
(HOES/LUES)-- are depicted in Fig.~\ref{fig:WFS}. From those plots
we observe that the electrons responsible for metallicity are
either (a) extended over both the SWNT and the silicon slab; (b)
predominantly over the silicon slab or (c) confined to the SWNT.
We also found an instance --the (12,4) SWNT in the parallel
configuration-- in which the wavefunction is \emph{localized} at
the interface between the nanotube and the surface. In most cases
involving the semiconducting nanotubes a more pronounced
contribution from the nanotube to the HOES is seen in contrast
with a more surface-like LUES. Those findings can be understood in
terms of the relative contributions to the band structure from the
carbon and silicon orbitals as seen from the PDOS in
Fig.~\ref{fig:Figbands}.

\subsection{Adsorption energies}
\begin{figure}
\scalebox{0.4}{\includegraphics{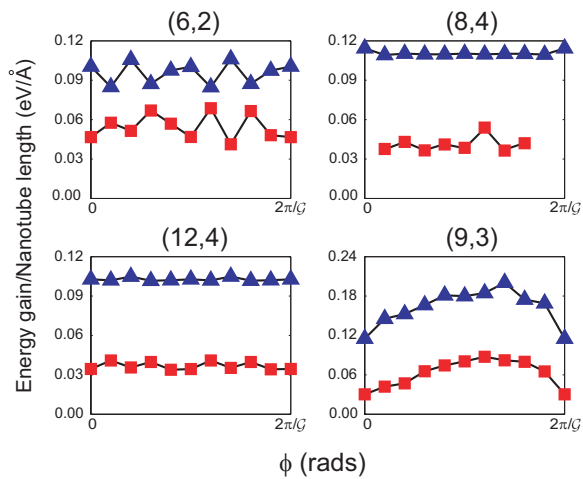}}
\caption{\label{fig:Fig4}(Color online) Energy gain vs. axial
rotation, prior to the relaxation cycle and at fixed height. Red
squares show results in the parallel configuration, while the blue
triangles correspond to the perpendicular configuration.
$\mathcal{G}\equiv{}\text{GCD}(m,n)$. The results shown here help
identify the best angular configuration for chiral nanotubes,
which can not be known \emph{a priori}. The largest adsorption
energies occur for tubes parallel to the trench. The energy
dependence on angle of rotation is more marked as the nanotube
diameter is decreased, since is brought in closer proximity with
the surface. The metallic nanotube shows the most pronounced
energy dependence.}
\end{figure}

Adsorption energies on the Si(100) surface for the carbon
nanotubes in this study are presented in Table \ref{tab:tableAE}.
 The adsorption energies are obtained as the difference
between the total energies in the relaxed hybrid structures and
those of the fully relaxed tubes and surface in the same supercell
separated by 10~\AA. Notice the constant adsorption energy for
semiconducting nanotubes in the perpendicular configuration, and
the decreasing adsorption energy with increasing diameter for
semiconducting tubes in the parallel configuration. These two
trends meet for a tube with diameter in between the (8,4) and
(12,4) nanotubes, after which a semiconducting tube in the
perpendicular configuration has a stronger adsorption energy. This
is stressed by the bold font in Table \ref{tab:tableAE}.
\begin{table}
\caption{\label{tab:tableAE} Adsorption energies per unit length
(eV/\AA) after full relaxation has been achieved. Notice the lower
adsorption energies obtained for semiconducting nanotubes. For the
(12,4) tube (in bold) the perpendicular configuration turns out to
be the most favorable.}
\begin{ruledtabular}
\begin{tabular}{lrrrr}
Nanotube:&(6,2)&(8,4)&(12,4)&(9,3)\\
\hline
Parallel&0.20 &0.18 &\bf{0.11} &0.63\\
Perpendicular&0.15 &0.15 &\bf{0.15} &0.60
\end{tabular}
\end{ruledtabular}
\end{table}
 In order to determine the best angular `starting' structure a chiral nanotube has to be rotated
about its axis. Chirality precludes a highly symmetric
configuration, potentially with the lowest energy, from being
easily visualized. The energies as a function of $\phi$ with no
atomic relaxation were obtained, and the difference of those
energies with respect to the energy of a system where the tubes
and surface are 10 \AA{} apart are reported (this is called the
energy gain and it is not an adsorption energy as the structures
are not fully relaxed). The results are shown in
Fig.~\ref{fig:Fig4}, where the lower curves (red squares) refer to
SWNTs in the perpendicular configuration. The upper curves (blue
triangles) refer to the SWNTs in the parallel configuration. For
our chiral tubes, the range of angles is given by
 $(0,2\pi/\text{GCD}(m,n))$, where GCD$(m,n)$ is the greatest common
 divisor of $m$ and $n$. This range gives the maximum angular
 freedom that the nanotubes will have before the position of their
 carbon atoms becomes periodic. Notice the almost
complete independence on $\phi$ for this energy for all
semiconducting tubes, due to their weaker bonding as compared with
metallic nanotubes. The (6,2) nanotube with the smallest studied
diameter shows the most pronounced
 oscillations. This is due to its close proximity to the Si(100)
 surface that can greatly modify its energy as carbon and silicon atoms
 are brought closer to each other as a result of the tube's
 rotation about its own axis. The distinctive trend for the metallic
tube indicates an optimal configuration in which an extra bond
(for a total of eight bonds) might form when full relaxation is
performed.  All the previous results shown in this paper were
obtained at $\phi=0.0$. Despite the fact that the metallic tube
might not be in optimal angular configuration, our results are the
first to date to show distinctively different trends for
semiconducting and metallic nanotubes. In fact, once the metallic
tube is in its optimal configuration, the trends provided here
will only be accentuated.
In the previously studied cases
 (Refs.~\onlinecite{Orellana2004_1,Orellana2004_2,Orellana05}),
the highly symmetric atomic arrangement of the achiral (6,6) SWNT
with respect to the underlying surface silicon atoms results in
optimal angular configurations in which $D$ is a function of
$\phi$. We consider that in a more general setting, chirality
precludes such a highly symmetric configuration from occurring:
the relative position of carbon atoms closest to the surface with
respect to surface silicon atoms becomes a complicated function of
the chiral angle. In Fig.~\ref{fig:Fig4} we observe a small energy
dependence per unit length (of the order of 100 meV/\AA) against
the tube's angle of rotation $\phi$. This justifies the angular
sweeping at fixed $D$ for our chiral SWNTs.

 It is also apparent from Fig.~\ref{fig:Fig4} that the metallic
nanotube displays an energy dependence on the angle of rotation
twice as large as the semiconducting SWNTs. This is consistent
with the fact that metallic tubes stay closer to the surface in
comparison to semiconducting nanotubes of the same diameter. This
could also be influenced by the different surface reconstruction
we employed.

The prominent difference in the adsorption energies presented in
Table \ref{tab:tableAE} suggests a mechanism for the separation
between semiconducting and metallic nanotubes in solid phase over
this diameter range, provided they are always adsorbed in
configurations with the greatest energy gain. Experimental
techniques for nanotube separation to date rely on the chemistry
and dipole moments of samples \emph{in
solution}.\cite{Krupke2003,Strano2003}

\section{conclusions}\label{sec:conclusions}
 We have studied semiconducting SWNTs adsorbed on the Si(100)
surface. We used the results from the (9,3) SWNT, which compare
well to previous published results for metallic
SWNTs\cite{Orellana2004_1,Orellana2004_2} as a test bed for our
choice of exchange-correlation potential and surface
reconstruction. Semiconducting SWNTs of the diameter range studied
are placed at an almost constant distance to the surface. Those
tubes are 0.5 \AA{} closer to the surface when they are above and
aligned with the trench between adjacent dimer rows, in comparison
with any other configuration in which the trench and the tube axis
do not align. We found a weak angular dependence on the adsorption
energy of the system, and we believe this dependence will be
further lowered as the length of the nanotube's unit cell
increases, due to the loss of symmetry in the relative positions
of the carbon atoms in the SWNT closest to the Si(100) surface
atoms. For the (9,3) SWNT, we found in agreement with
Refs.~\onlinecite{Orellana2004_1,Orellana2004_2} that it remains
metallic in either configuration, but the weak angular dependence
tells us it will remain at an almost fixed height $D$, independent
of its angular orientation. We also found a smaller adsorption
energy for semiconducting SWNTs in comparison to metallic SWNTs of
similar diameter, consistent with weaker bonding.

 The electronic properties of
these hybrid systems will vary in a qualitative way according to
the relative orientation of the SWNT (parallel, perpendicular)
with respect to the surface, but we find from our calculations
that the system composed of semiconducting tubes on the
semiconducting Si(100) surface displays dramatically smaller gaps
in comparison with the isolated semiconducting systems that
compose it, in both studied configurations. This reduction of the
semiconducting gap can be ascribed to the modification of the band
structure due to the surface reconfiguration, electronic charge
transfer from the nanotubes and the resulting band hybridization
from carbon states at energies where the gap appears for isolated
tubes, as the PDOS indicates. The HOES for the hybrid systems can
be extended, located predominately over either the
 silicon slab or the SWNT, or be localized
 at the interface between the SWNT and the Si(100) forming a one-dimensional conduction
channel. In contrast, the LUES tend to be more localized towards
the surface when semiconducting nanotubes are brought into
proximity. Currently, experimental results on this system are
starting to emerge.\cite{Albrechtu} We hope that the results
provided in this paper motivate further experimental work in the
area, as the properties described in here might be useful for
electronic as well as opto-electronic applications.

Systematic studies as the one presented here are truly necessary
in order to assess robust properties of this hybrid system against
properties that might appear for a given choice of chiral angles
and geometrical configuration. In our case, the gap reduction
appears for both configurations, and the bonding lengths are found
to be consistently larger for semiconducting nanotubes in
comparison with metallic ones. This implies a reduced absorption
energy for semiconducting nanotubes on this surface. In turn, this
might serve to mechanically attach metallic nanotubes and release
semiconducting ones by suitable heating under dry
conditions.\cite{Krupke2003,Strano2003}

\section*{ACKNOWLEDGMENTS}
 G. Bauer and G. Lopez-Walle assisted during the early
stages of this project. We thank R. M. Martin, N. Tayebi, K.
Ritter, H. Terrones and J. Junquera for useful discussions. M.
Pruneda assisted us in performing the Voroni and Hirshfield charge
analysis. These additional SIESTA routines were written by P.
Ordej\'on. Calculations were performed on the CSE mac OS X Turing
cluster, linux CEG cluster (U. Ravaioli, Z. Yang). This work was
supported by National Computational Science Alliance under grant
number DMR050032N (NCSA's IBM pSeries 690 Copper cluster) and the
Office of Naval Research, grant N00014-98-1-0604. S. B.-L.
acknowledges partial funding from CONACyT (Mexico), P.M.A. a NDSEG
graduate fellowship and N.A.R. grants from NSF (DMR-0325939) and
DoE (DEFG02-91ER45439).


\begin{thebibliography}{29}
\expandafter\ifx\csname
natexlab\endcsname\relax\def\natexlab#1{#1}\fi
\expandafter\ifx\csname bibnamefont\endcsname\relax
  \def\bibnamefont#1{#1}\fi
\expandafter\ifx\csname bibfnamefont\endcsname\relax
  \def\bibfnamefont#1{#1}\fi
\expandafter\ifx\csname citenamefont\endcsname\relax
  \def\citenamefont#1{#1}\fi
\expandafter\ifx\csname url\endcsname\relax
  \def\url#1{\texttt{#1}}\fi
\expandafter\ifx\csname
urlprefix\endcsname\relax\def\urlprefix{URL }\fi
\providecommand{\bibinfo}[2]{#2}
\providecommand{\eprint}[2][]{\url{#2}}

\bibitem[{\citenamefont{Kim et~al.}(2004{\natexlab{a}})\citenamefont{Kim,
  Leben, and Zhang}}]{Kim2004}
\bibinfo{author}{\bibfnamefont{Y.-H.} \bibnamefont{Kim}},
  \bibinfo{author}{\bibfnamefont{M.~J.} \bibnamefont{Leben}}, \bibnamefont{and}
  \bibinfo{author}{\bibfnamefont{S.~B.} \bibnamefont{Zhang}},
  \bibinfo{journal}{Phys. Rev. Lett.} \textbf{\bibinfo{volume}{92}},
  \bibinfo{pages}{176102} (\bibinfo{year}{2004}{\natexlab{a}}).

\bibitem[{\citenamefont{Kim et~al.}(2004{\natexlab{b}})\citenamefont{Kim,
  Heben, and Zhang}}]{Kim2005}
\bibinfo{author}{\bibfnamefont{Y.-H.} \bibnamefont{Kim}},
  \bibinfo{author}{\bibfnamefont{M.~J.} \bibnamefont{Heben}}, \bibnamefont{and}
  \bibinfo{author}{\bibfnamefont{S.~B.} \bibnamefont{Zhang}},
  \bibinfo{journal}{AIP Conf. Proc.} \textbf{\bibinfo{volume}{772}},
  \bibinfo{pages}{1031} (\bibinfo{year}{2004}{\natexlab{b}}).

\bibitem[{\citenamefont{Orellana et~al.}(2003)\citenamefont{Orellana, Miwa, and
  Fazzio}}]{Orellana2004_1}
\bibinfo{author}{\bibfnamefont{W.}~\bibnamefont{Orellana}},
  \bibinfo{author}{\bibfnamefont{R.~H.} \bibnamefont{Miwa}}, \bibnamefont{and}
  \bibinfo{author}{\bibfnamefont{A.}~\bibnamefont{Fazzio}},
  \bibinfo{journal}{Phys.\ Rev.\ Lett.} \textbf{\bibinfo{volume}{91}},
  \bibinfo{pages}{166802} (\bibinfo{year}{2003}).

\bibitem[{\citenamefont{Orellana et~al.}(2004)\citenamefont{Orellana, Miwa, and
  Fazzio}}]{Orellana2004_2}
\bibinfo{author}{\bibfnamefont{W.}~\bibnamefont{Orellana}},
  \bibinfo{author}{\bibfnamefont{R.~H.} \bibnamefont{Miwa}}, \bibnamefont{and}
  \bibinfo{author}{\bibfnamefont{A.}~\bibnamefont{Fazzio}},
  \bibinfo{journal}{Surf.\ Sci.} \textbf{\bibinfo{volume}{566-568}},
  \bibinfo{pages}{728} (\bibinfo{year}{2004}).

\bibitem[{\citenamefont{Miwa et~al.}(2005)\citenamefont{Miwa, Orellana, and
  Fazzio}}]{Orellana05}
\bibinfo{author}{\bibfnamefont{R.~H.} \bibnamefont{Miwa}},
  \bibinfo{author}{\bibfnamefont{W.}~\bibnamefont{Orellana}}, \bibnamefont{and}
  \bibinfo{author}{\bibfnamefont{A.}~\bibnamefont{Fazzio}},
  \bibinfo{journal}{Appl. Phys. Lett.} \textbf{\bibinfo{volume}{86}},
  \bibinfo{pages}{213111} (\bibinfo{year}{2005}).

\bibitem[{\citenamefont{Albrecht and Lyding}(2003)}]{Albrecht2003}
\bibinfo{author}{\bibfnamefont{P.~M.} \bibnamefont{Albrecht}} \bibnamefont{and}
  \bibinfo{author}{\bibfnamefont{J.~W.} \bibnamefont{Lyding}},
  \bibinfo{journal}{Appl. Phys. Lett.} \textbf{\bibinfo{volume}{83}},
  \bibinfo{pages}{5029} (\bibinfo{year}{2003}).

\bibitem[{\citenamefont{Albrecht and Lyding}(2004)}]{Albrecht2004}
\bibinfo{author}{\bibfnamefont{P.~M.} \bibnamefont{Albrecht}} \bibnamefont{and}
  \bibinfo{author}{\bibfnamefont{J.~W.} \bibnamefont{Lyding}},
  \bibinfo{journal}{AIP Conf. Proc.} \textbf{\bibinfo{volume}{723}},
  \bibinfo{pages}{173} (\bibinfo{year}{2004}).

\bibitem[{\citenamefont{Ruppalt et~al.}(2004)\citenamefont{Ruppalt, Albrecht,
  and Lyding}}]{Ruppalt2004}
\bibinfo{author}{\bibfnamefont{L.~B.} \bibnamefont{Ruppalt}},
  \bibinfo{author}{\bibfnamefont{P.~M.} \bibnamefont{Albrecht}},
  \bibnamefont{and} \bibinfo{author}{\bibfnamefont{J.~W.}
  \bibnamefont{Lyding}}, \bibinfo{journal}{J. Vac. Sci. Technol. B}
  \textbf{\bibinfo{volume}{22}}, \bibinfo{pages}{1071} (\bibinfo{year}{2004}).

\bibitem[{\citenamefont{Tzolov et~al.}(2004)\citenamefont{Tzolov, Chang, Yin,
  Straus, and Xu}}]{Tzolov2004}
\bibinfo{author}{\bibfnamefont{M.}~\bibnamefont{Tzolov}},
  \bibinfo{author}{\bibfnamefont{B.}~\bibnamefont{Chang}},
  \bibinfo{author}{\bibfnamefont{A.}~\bibnamefont{Yin}},
  \bibinfo{author}{\bibfnamefont{D.}~\bibnamefont{Straus}}, \bibnamefont{and}
  \bibinfo{author}{\bibfnamefont{J.~M.} \bibnamefont{Xu}},
  \bibinfo{journal}{Phys. Rev. Lett.} \textbf{\bibinfo{volume}{92}},
  \bibinfo{pages}{075505} (\bibinfo{year}{2004}).

\bibitem[{\citenamefont{Jensen et~al.}(2004)\citenamefont{Jensen, Hauptmann,
  Nygard, Sadowski, and Lindelof}}]{Jensen2004}
\bibinfo{author}{\bibfnamefont{A.}~\bibnamefont{Jensen}},
  \bibinfo{author}{\bibfnamefont{J.~R.} \bibnamefont{Hauptmann}},
  \bibinfo{author}{\bibfnamefont{J.}~\bibnamefont{Nygard}},
  \bibinfo{author}{\bibfnamefont{J.}~\bibnamefont{Sadowski}}, \bibnamefont{and}
  \bibinfo{author}{\bibfnamefont{P.~E.} \bibnamefont{Lindelof}},
  \bibinfo{journal}{Nano Lett.} \textbf{\bibinfo{volume}{4}},
  \bibinfo{pages}{349} (\bibinfo{year}{2004}).

\bibitem[{\citenamefont{Su et~al.}(2000)\citenamefont{Su, Li, Maynor, Buldum,
  Lu, and Liu}}]{Su2000}
\bibinfo{author}{\bibfnamefont{M.}~\bibnamefont{Su}},
  \bibinfo{author}{\bibfnamefont{Y.}~\bibnamefont{Li}},
  \bibinfo{author}{\bibfnamefont{B.}~\bibnamefont{Maynor}},
  \bibinfo{author}{\bibfnamefont{A.}~\bibnamefont{Buldum}},
  \bibinfo{author}{\bibfnamefont{J.~P.} \bibnamefont{Lu}}, \bibnamefont{and}
  \bibinfo{author}{\bibfnamefont{J.}~\bibnamefont{Liu}}, \bibinfo{journal}{J.
  Phys. Chem. B} \textbf{\bibinfo{volume}{104}}, \bibinfo{pages}{6505}
  (\bibinfo{year}{2000}).

\bibitem[{\citenamefont{Hohenberg and Kohn}(1964)}]{hohenberg64}
\bibinfo{author}{\bibfnamefont{P.}~\bibnamefont{Hohenberg}} \bibnamefont{and}
  \bibinfo{author}{\bibfnamefont{W.}~\bibnamefont{Kohn}},
  \bibinfo{journal}{Phys. Rev.} \textbf{\bibinfo{volume}{136}},
  \bibinfo{pages}{B864} (\bibinfo{year}{1964}).

\bibitem[{\citenamefont{Fonseca-Guerra
  et~al.}(2003)\citenamefont{Fonseca-Guerra, Handgraaf, Baerends, and
  Bickelhaupt}}]{Fonseca2003}
\bibinfo{author}{\bibfnamefont{C.}~\bibnamefont{Fonseca-Guerra}},
  \bibinfo{author}{\bibfnamefont{J.-W.} \bibnamefont{Handgraaf}},
  \bibinfo{author}{\bibfnamefont{E.~J.} \bibnamefont{Baerends}},
  \bibnamefont{and} \bibinfo{author}{\bibfnamefont{F.~M.}
  \bibnamefont{Bickelhaupt}}, \bibinfo{journal}{J. Comput. Chem.}
  \textbf{\bibinfo{volume}{25}}, \bibinfo{pages}{189} (\bibinfo{year}{2003}).

\bibitem[{\citenamefont{Kohn and Sham}(1965)}]{Kohn65}
\bibinfo{author}{\bibfnamefont{W.}~\bibnamefont{Kohn}} \bibnamefont{and}
  \bibinfo{author}{\bibfnamefont{L.~J.} \bibnamefont{Sham}},
  \bibinfo{journal}{Phys. Rev.} \textbf{\bibinfo{volume}{140}},
  \bibinfo{pages}{A1133} (\bibinfo{year}{1965}).

\bibitem[{\citenamefont{Soler et~al.}(2002)\citenamefont{Soler, Artacho, Gale,
  Garcia, Junquera, Ordejon, and Sanchez-Portal}}]{Soler2002}
\bibinfo{author}{\bibfnamefont{J.~M.} \bibnamefont{Soler}},
  \bibinfo{author}{\bibfnamefont{E.}~\bibnamefont{Artacho}},
  \bibinfo{author}{\bibfnamefont{J.~D.} \bibnamefont{Gale}},
  \bibinfo{author}{\bibfnamefont{A.}~\bibnamefont{Garcia}},
  \bibinfo{author}{\bibfnamefont{J.}~\bibnamefont{Junquera}},
  \bibinfo{author}{\bibfnamefont{P.}~\bibnamefont{Ordejon}}, \bibnamefont{and}
  \bibinfo{author}{\bibfnamefont{D.}~\bibnamefont{Sanchez-Portal}},
  \bibinfo{journal}{J. Phys.: Condens. Matter} \textbf{\bibinfo{volume}{14}},
  \bibinfo{pages}{2745} (\bibinfo{year}{2002}).

\bibitem[{\citenamefont{Perdew and Zunger}(1981)}]{pz}
\bibinfo{author}{\bibfnamefont{J.~P.} \bibnamefont{Perdew}} \bibnamefont{and}
  \bibinfo{author}{\bibfnamefont{A.}~\bibnamefont{Zunger}},
  \bibinfo{journal}{Phys. Rev. B} \textbf{\bibinfo{volume}{23}},
  \bibinfo{pages}{5048} (\bibinfo{year}{1981}).

\bibitem[{\citenamefont{Ceperley and Alder}(1981)}]{ca}
\bibinfo{author}{\bibfnamefont{D.~M.} \bibnamefont{Ceperley}} \bibnamefont{and}
  \bibinfo{author}{\bibfnamefont{B.~J.} \bibnamefont{Alder}},
  \bibinfo{journal}{Phys. Rev. Lett.} \textbf{\bibinfo{volume}{45}},
  \bibinfo{pages}{566} (\bibinfo{year}{1981}).

\bibitem[{\citenamefont{Troullier and Martins}(1991)}]{tm}
\bibinfo{author}{\bibfnamefont{N.}~\bibnamefont{Troullier}} \bibnamefont{and}
  \bibinfo{author}{\bibfnamefont{J.~L.} \bibnamefont{Martins}},
  \bibinfo{journal}{Phys. Rev. B} \textbf{\bibinfo{volume}{43}},
  \bibinfo{pages}{1993} (\bibinfo{year}{1991}).

\bibitem[{\citenamefont{Junquera et~al.}(2001)\citenamefont{Junquera, Paz,
  Sanchez-Portal, and Artacho}}]{Junquera01}
\bibinfo{author}{\bibfnamefont{J.}~\bibnamefont{Junquera}},
  \bibinfo{author}{\bibfnamefont{O.}~\bibnamefont{Paz}},
  \bibinfo{author}{\bibfnamefont{D.}~\bibnamefont{Sanchez-Portal}},
  \bibnamefont{and} \bibinfo{author}{\bibfnamefont{E.}~\bibnamefont{Artacho}},
  \bibinfo{journal}{Phys. Rev. B} \textbf{\bibinfo{volume}{64}},
  \bibinfo{pages}{235111} (\bibinfo{year}{2001}).

\bibitem[{\citenamefont{Press et~al.}(1992)\citenamefont{Press, Teukolsky,
  Vetterling, and Flannery}}]{Pressrecipes}
\bibinfo{author}{\bibfnamefont{W.~H.} \bibnamefont{Press}},
  \bibinfo{author}{\bibfnamefont{S.~A.} \bibnamefont{Teukolsky}},
  \bibinfo{author}{\bibfnamefont{W.~T.} \bibnamefont{Vetterling}},
  \bibnamefont{and} \bibinfo{author}{\bibfnamefont{B.~P.}
  \bibnamefont{Flannery}}, \emph{\bibinfo{title}{Numerical recipes in fortran
  77}}, vol.~\bibinfo{volume}{1} of \emph{\bibinfo{series}{Fortran Numerical
  Recipes}} (\bibinfo{publisher}{Cambridge U. Press}, \bibinfo{year}{1992}),
  \bibinfo{edition}{2nd} ed.

\bibitem[{\citenamefont{Aschcroft and Mermin}(1976)}]{AshcroftMermin}
\bibinfo{author}{\bibfnamefont{N.~W.} \bibnamefont{Aschcroft}}
  \bibnamefont{and} \bibinfo{author}{\bibfnamefont{N.~D.}
  \bibnamefont{Mermin}}, \emph{\bibinfo{title}{Solid State Physics}}
  (\bibinfo{publisher}{Harcourt College Publishing}, \bibinfo{address}{Fort
  Worth}, \bibinfo{year}{1976}), p.~\bibinfo{pages}{76}.

\bibitem[{\citenamefont{Kganyago and Ngoepe}(2003)}]{Kganyago2003}
\bibinfo{author}{\bibfnamefont{K.~R.} \bibnamefont{Kganyago}} \bibnamefont{and}
  \bibinfo{author}{\bibfnamefont{P.~E.} \bibnamefont{Ngoepe}},
  \bibinfo{journal}{Phys. Rev. B} \textbf{\bibinfo{volume}{68}},
  \bibinfo{pages}{205111} (\bibinfo{year}{2003}).

\bibitem[{\citenamefont{Healy et~al.}(2001)\citenamefont{Healy, Filippi,
  Kratzer, Penev, and Scheffler}}]{Healy2001}
\bibinfo{author}{\bibfnamefont{S.~B.} \bibnamefont{Healy}},
  \bibinfo{author}{\bibfnamefont{C.}~\bibnamefont{Filippi}},
  \bibinfo{author}{\bibfnamefont{P.}~\bibnamefont{Kratzer}},
  \bibinfo{author}{\bibfnamefont{E.}~\bibnamefont{Penev}}, \bibnamefont{and}
  \bibinfo{author}{\bibfnamefont{M.}~\bibnamefont{Scheffler}},
  \bibinfo{journal}{Phys. Rev. Lett.} \textbf{\bibinfo{volume}{87}},
  \bibinfo{pages}{016105} (\bibinfo{year}{2001}).

\bibitem[{\citenamefont{Monkhorst and Pack}(1976)}]{mp76}
\bibinfo{author}{\bibfnamefont{H.~J.} \bibnamefont{Monkhorst}}
  \bibnamefont{and} \bibinfo{author}{\bibfnamefont{J.~D.} \bibnamefont{Pack}},
  \bibinfo{journal}{Phys. Rev. B} \textbf{\bibinfo{volume}{13}},
  \bibinfo{pages}{5188} (\bibinfo{year}{1976}).

\bibitem[{\citenamefont{Hertel et~al.}(1998)\citenamefont{Hertel, Walkup, and
  Avouris}}]{Hertel1998}
\bibinfo{author}{\bibfnamefont{T.}~\bibnamefont{Hertel}},
  \bibinfo{author}{\bibfnamefont{R.~E.} \bibnamefont{Walkup}},
  \bibnamefont{and} \bibinfo{author}{\bibfnamefont{P.}~\bibnamefont{Avouris}},
  \bibinfo{journal}{Phys. Rev. B} \textbf{\bibinfo{volume}{58}},
  \bibinfo{pages}{13870} (\bibinfo{year}{1998}).

\bibitem[{\citenamefont{Albrecht and Lyding}(2005)}]{Albrechtu}
\bibinfo{author}{\bibfnamefont{P.~M.} \bibnamefont{Albrecht}} \bibnamefont{and}
  \bibinfo{author}{\bibfnamefont{J.~W.} \bibnamefont{Lyding}},
  \bibinfo{journal}{unpublished}  (\bibinfo{year}{2005}).

\bibitem[{\citenamefont{Berber and Oshiyama}(2006)}]{Berber2006}
\bibinfo{author}{\bibfnamefont{S.}~\bibnamefont{Berber}} \bibnamefont{and}
  \bibinfo{author}{\bibfnamefont{A.}~\bibnamefont{Oshiyama}},
  \bibinfo{journal}{Phys. Rev. Lett.} \textbf{\bibinfo{volume}{96}},
  \bibinfo{pages}{105505} (\bibinfo{year}{2006}).

\bibitem[{\citenamefont{Krupke et~al.}(2003)\citenamefont{Krupke, Hennrich,
  v.~L\"hneysen, and Kappes}}]{Krupke2003}
\bibinfo{author}{\bibfnamefont{R.}~\bibnamefont{Krupke}},
  \bibinfo{author}{\bibfnamefont{F.}~\bibnamefont{Hennrich}},
  \bibinfo{author}{\bibfnamefont{H.}~\bibnamefont{v.~L\"hneysen}},
  \bibnamefont{and} \bibinfo{author}{\bibfnamefont{M.~M.}
  \bibnamefont{Kappes}}, \bibinfo{journal}{Science}
  \textbf{\bibinfo{volume}{301}}, \bibinfo{pages}{344} (\bibinfo{year}{2003}).

\bibitem[{\citenamefont{Strano et~al.}(2003)\citenamefont{Strano, Dyke, Usrey,
  Barone, Allen, Shan, Kittrell, Hauge, Tour, and Smalley}}]{Strano2003}
\bibinfo{author}{\bibfnamefont{M.~S.} \bibnamefont{Strano}},
  \bibinfo{author}{\bibfnamefont{C.~A.} \bibnamefont{Dyke}},
  \bibinfo{author}{\bibfnamefont{M.~L.} \bibnamefont{Usrey}},
  \bibinfo{author}{\bibfnamefont{P.~W.} \bibnamefont{Barone}},
  \bibinfo{author}{\bibfnamefont{M.~J.} \bibnamefont{Allen}},
  \bibinfo{author}{\bibfnamefont{H.}~\bibnamefont{Shan}},
  \bibinfo{author}{\bibfnamefont{C.}~\bibnamefont{Kittrell}},
  \bibinfo{author}{\bibfnamefont{R.~H.} \bibnamefont{Hauge}},
  \bibinfo{author}{\bibfnamefont{J.~M.} \bibnamefont{Tour}}, \bibnamefont{and}
  \bibinfo{author}{\bibfnamefont{R.~E.} \bibnamefont{Smalley}},
  \bibinfo{journal}{Science} \textbf{\bibinfo{volume}{301}},
  \bibinfo{pages}{1519} (\bibinfo{year}{2003}).

\end{thebibliography}
\end{document}